\documentclass[letter,twocolumn,showpacs,aps,prb,floatfix]{revtex4}
\usepackage{graphicx,subfigure,epsfig,verbatim,psfrag,amsmath,amssymb,color}
\usepackage{indentfirst}
\usepackage{textcomp}
\usepackage{latexsym}
\usepackage{amssymb}
\usepackage{amsmath}
\makeatletter
\newcommand{\mean}[1]{\left<#1\right>}
\begin{document}

\title{Thermodynamics of Ising Spins on the Star Lattice}
\author{Zewei Chen}
\author{Nvsen Ma}
\author{Dao-Xin Yao}
\email{Email: yaodaox@mail.sysu.edu.cn}

\affiliation{State Key Laboratory of Optoelectronic Materials and
Technologies, School of Physics and Engineering, Sun Yat-sen
University, Guangzhou 510275, China}

\begin{abstract}
There is a new class of two-dimensional magnetic materials polymeric iron (III) acetate fabricated recently in which Fe ions form a star lattice. We study the thermodynamics of Ising spins on the star lattice with exact analytic method and Monte Carlo simulations. Mapping the star lattice to the honeycomb lattice, we
obtain the partition function for the system with asymmetric interactions. The free energy, internal energy, specific heat, entropy and susceptibility are presented, which can be used to determine the sign of the interactions in the real materials. Moreover, we find the rich phase diagrams of the system as a
function of interactions, temperature and external magnetic field. For frustrated interactions without external field, the ground state is disordered (spin liquid) with residual entropy $1.522\ldots$ per unit cell. When a weak field is applied, the system enters a ferrimagnetic phase with residual entropy $ln4$ per unit cell.
\end{abstract}
\pacs{75.10.Hk, 75.30.Kz, 75.40.-s, 64.60.-i}
\maketitle
%@@@@@@@@@@@@@@@@@@@@@@@@@@@@@@@@@@@@@@@@@@@@@@@@@@@@@@@@@@@@@@@@@@@@@@@@@@@@@@@

\section{Introduction}
Spin systems with geometrical frustration have both fundamental and practical importance. Theoretically, lots of interesting phenomena have been found in the geometrically frustrated systems, like the antiferromagnetic triangular lattice, kagome lattice. The systems can remain disordered even at absolute zero temperature because of the competitive magnetic interactions. For example, an antiferromagnetic triangular lattice has a residual entropy $s_0=0.3281\ldots$ per unit cell~\cite{PhysRev.79.357}. The frustration effect has important application in achieving a lower temperatures through the adiabatic demagnetization compared with other methods. When the temperature, external magnetic field, and other factors are considered, the geometrically frustrated systems can show very rich phase diagrams.~\cite{diep} A typical case is that the magnetocaloric effect can be enhanced near the phase
transition points when a finite external field is applied.~\cite{2004PhRvB70j0403Z,2004PhRvB70j4418I,JPSJ.73.2851} Of considerable interest has been searching for geometrically frustrated systems. Some new frustrated materials have been fabricated and studied recently, such as $Ho_2Ti_2O_7$, $Dy_2Ti_2O_7$, and $Cu_9X_2(cpa)_6\cdot xH_2O (X=F, Cl, Br)$.~\cite{2000JAP87.5914R,1997PhRvL79.2554H, 1994JAP75.5949M,tlk,ISI:000255457200055,2008PhRvB78v4410L,2008PhRvB78b4428Y}

\begin{figure}[!t]
\centering
{\resizebox*{0.47\textwidth}{!}{\includegraphics{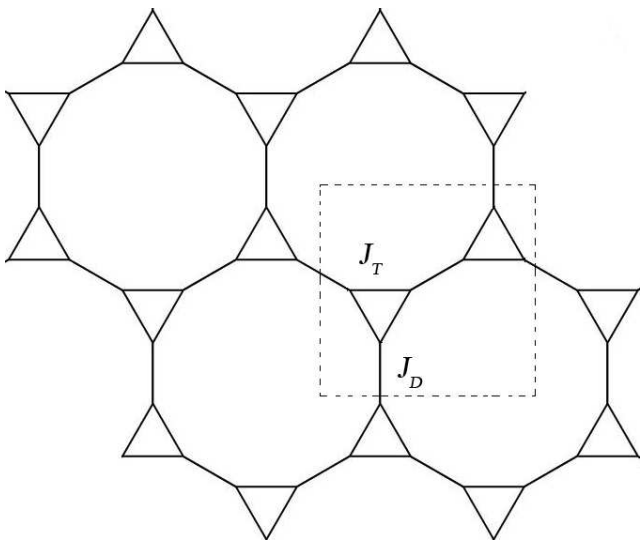}}}
\caption{(Color online). The star lattice. The dashed frame represents a unit cell of the star lattice. There are six spins per unit cell.} %, six $J_{T}$ bonds and one $J_{D}$ bond.}
\label{starlattice}
\end{figure}

In 2007, a new class of geometrically frustrated magnetic materials polymeric iron (III) acetate ~\cite{17615608} was fabricated, in which Fe ions form a two-dimensional lattice referred as star lattice. Experiment has found that the materials exhibit spin frustration and have two kinds of magnetic interactions: intratrimer $J_T$ and intertrimer $J_D$ shown in Fig.~\ref{starlattice}. The Fe ion has a large spin, which is $S=5/2$. The system may have the paramagnetic ground state because of the geometrical and quantum fluctuations.

The Ising model on the star lattice with uniform ferromagnetic couplings has been solved.~\cite{1995PhRvB..51.5840B} The critical temperature $K_{c}=0.81201$ was given with $K_{c}=\beta_{c}J$. The Bose-Hubbard model on the star lattice has also been studied using the quantum Monte Carlo method.~\cite{2009PhRvB..80u4503I} Recently, the edge states and topological orders were found in the spin liquid phases of star lattice.~\cite{1202.4163}  Even though the $S=1/2$ quantum Heisenberg model is considered to be appropriate to help studying the new material because of its quantum fluctuations in the ground state,~\cite{springerlink:10.1007/BFb0119592} the Ising model on the star lattice is still very important especially for the non-uniform case. In real materials, the Fe ion has $S=5/2$ which is close to the classical limit and the magnetic system shows two types of interactions.~\cite{17615608} Therefore, it is important to study Ising spins on the star lattice with the asymmetric interactions, especially for the frustrated case.

In this paper, we aim at the thermodynamics of Ising spins on the star lattice with asymmetric interactions using the exact analytic methods and Monte Carlo simulations. We present the phase diagram as a function of interactions, temperature and external magnetic field. There is a clear difference between the $J_{D}>0$ case and the $J_{D}<0$ case. Our study provides useful information for determining the sign of $J_{D}$.

This paper is organized as follows. The model is described in Sec. II. In Sec. III, we map the star lattice to honeycomb lattice and get the exact results. Sec. IV presents the phase diagrams as functions of interactions and external magnetic field. In Sec. V, the Monte Carlo results for the heat capacity and susceptibility are given. Finally, in Sec. VI we summarize the results.

%%%%%%% F I G U R E %%%%%%%%%%%%

%%%%%%% E N D   F I G U R E %%%%%%%%%%%%

%@@@@@@@@@@@@@@@@@@@@@@@@@@@@@@@@@@@@@@@@@@@@@@@@@@@@@@@@@@@@@@@@@@@@@@@@@@@@@@@
\section{Model \label{model}}
The structure of star lattice and its unit cell are illustrated in Fig.~\ref{starlattice}. We study Ising spins on the star lattice with two kinds of nearest-neighbor interactions, the intratrimer coupling $J_{T}$ and intertrimer coupling $J_{D}$. The Hamiltonian is
\begin{equation}
H=-J_{T}\sum_{<ij>}\sigma_{i}\sigma_{j}-J_{D}\sum_{<i'j'>}\sigma_{i'}\sigma_{j'}-h\sum_{i}\sigma_{i},
\end{equation}
where $<ij>$ runs over all the nearest neighbor spin pairs, $J_{T}$ is the intra-triangular interaction and $J_{D}$ is the inter-triangular interaction, $h$ is the external magnetic field, and $\sigma_{i}=\pm 1$. The unit cell of the star lattice contains six spins shown in Fig.~\ref{starlattice}. If we use $N_T$ and $N_D$ to denote the total numbers of $J_{T}$- and $J_D$-bonds, we have $N_T:N_D=2:1$. %six $J_{T}$ bonds and one $J_{D}$ bond.
The analytic result of partition function is obtained for $h=0$.

For simplicity, we use $|J_T|$ as the units of energy in the following. The corresponding phase diagrams are actually in a three-dimensional parameter space, $\frac{J_{D}}{|J_{T}|}$,$\frac{T}{|J_{T}|}$ and $\frac{h}{|J_{T}|}$. %The phase diagram is sensitive to the sign of $J_{T}$.

%%%%%%%%%%%%%%%%%%%%%%%%%%%%%%%%%%%%%%%%%%%%%%%%%%%%%%%%%%%%%%%%%%%%%%%%%%%%%%%
\section{Exact solution in zero field \label{zerofield}}
In this section, we study the exact analytic results of Ising model on the star lattice in zero magnetic field ($h=0$). Using a sequence of $\Delta-Y$ transformation and series reductions\cite{ISI:000255457200055},
we can transform the Ising model on the star lattice into one on the honeycomb lattice whose partition function has been exactly solved using the Pfaffian method~\cite{1966JMP7.1776F,1963JMP4}.
Besides the exact analytical results, we expand the partition function in series for some special cases.

\subsection{Effective coupling on the equivalent honeycomb lattice \label{sec:mapping}}
\begin{figure}[!t]
\centering
{\resizebox*{0.47\textwidth}{!}{\includegraphics{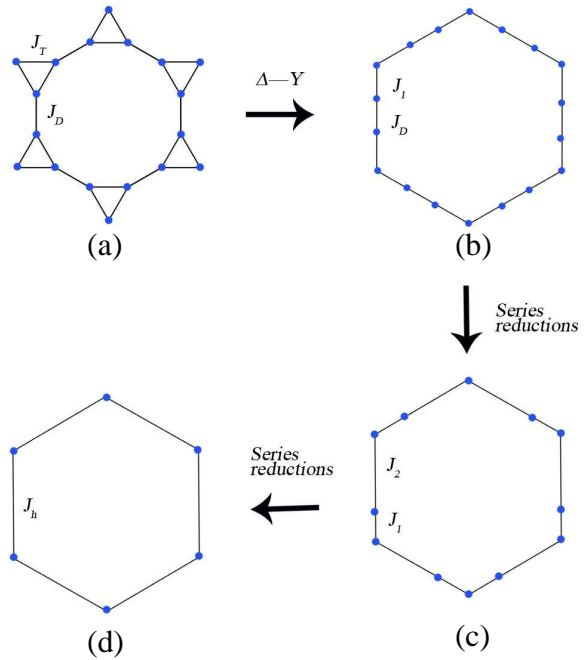}}}
\caption{(Color online). Transformation of star lattice to honeycomb lattice. (a) Depicts a section of the star lattice with two couplings, $J_{D}, J_{T}$. By applying the $\Delta-Y$ transformation and then we obtain (b). After that take the two bonds $J_{D}$, $J_{1}$ in series and we obtain (c) where generates a new coupling $J_{2}$. Finally, take $J_{1}$ and $J_{2}$ in series and the honeycomb lattice (d) is obtained.}
\label{transformation}
\end{figure}

The results of $\Delta-Y$ transformation and series reduction are given in Ref.~\onlinecite{ISI:000255457200055}. Using the variables $t_{i}=tanh\beta J_{i}$ and $x_{i}=e^{-2\beta J_{i}}$, the relations among the exchange couplings of Fig.~\ref{transformation} can be written as
\begin{align}
t_1  =           &\frac {1}{\sqrt{t_{T}+t_{T}^{-1}-1}}\\
& t_2   =t_{1}t_{D}\\
&t_{h} =t_{1}t_{2}.
\end{align}
We write $t_h$ in terms of $t_T$ and $t_D$ directly
\begin{equation}
t_h=\frac{t_{T}t_D}{t_{T}^2-t_{T}+1}
\end{equation}
For convenience, we can rewrite it in terms of $x_i$
\begin{equation}
x_h=\frac{x_D+(2+x_D)x_{T}^{2}}{1+(1+2x_D)x_{T}^2}
\label{xh}
\end{equation}

%@@@@@@@@@@@@@@@@@@@@@@@@@@@@@@@@@@@@@@@@@@@@@@@@@@@@@@@@@@@@@@@@@@@@@@@@@@@@@@@
%%%%%%% F I G U R E %%%%%%%%%%%%
%%%%%%% E N D   F I G U R E %%%%%%%%%%%%
\subsection{Phase boundary}
\begin{figure}[!t]
\centering
{\resizebox*{0.47\textwidth}{!}{\includegraphics{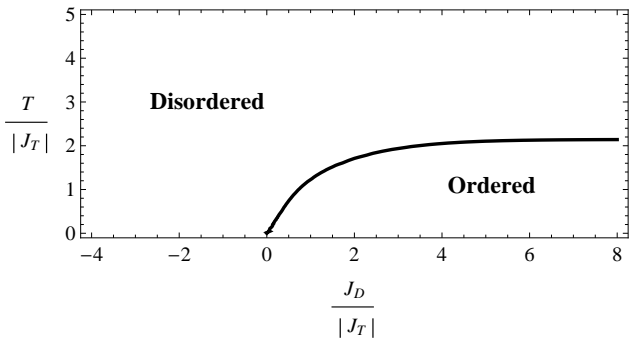}}}
\caption{(Color online). Phase diagram of the star lattice Ising model in the $(J_{D},T)$ plane, for $J_{T}= 1$ and $h=0$. The thick curve is the exact solution. It illustrates that when $J_{D}>0$, the phase is ordered.
The ordered phase is ferromagnetic. On the contrary, when $J_{D}<0$ the phase is immediately disordered (paramagnetic). When $J_{T}=-1$, the phase diagram is below $T$-axis (not shown in the figure), which implies that there is no phase transition in this phase.}
\label{zerofieldphasediagram}
\end{figure}

It is known that the critical temperature of the honeycomb lattice Ising model is given by $x_{h}^{c}=2-\sqrt{3}$. \cite{baxter} Having mapped star lattice to honeycomb lattice, we can substitute this into the equivalent coupling in Eq.~(\ref{xh}). Thus, an implicit equation for the critical temperature $\frac{1}{\beta_{c}}$ of star lattice Ising model can be obtained as,
\begin{equation}
\frac{e^{-2\beta_{c}J_{D}}+(2+e^{-2\beta_{c}J_{D}})e^{-4\beta_{c}J_{T}}}
{1+(1+2e^{-2\beta_{c}J_{D}})e^{-4\beta_{c}J_{T}}}=2-\sqrt{3}
\end{equation}
This result is plotted in Fig.~\ref{zerofieldphasediagram}. When $J_{D}$ is ferromagnetic ($J_{D}>0$) and strong enough, the critical temperature saturates at a finite value, i.e. $T_{c}/|J_{T}|\approx 4/ln[3/(2\sqrt{3}-3)]\approx 2.14332$. The critical temperature drops to zero as $J_{D}$ approaches zero. When $J_{T}\approx|J_{D}|$, the curve is approximately linear with $T_{c}/|J_{T}|\approx 1.23151$.

Furthermore, when $x_D=x_T=x$, Eq.(~\ref{xh}) reduces to the result of star lattice with equivalent couplings. \cite{1995PhRvB..51.5840B} We get $x_c=0.19710$, or equivalently,  $K_c=0.81201$.

Since all the factors obtained here are analytical, the singularity in the partition function remains when we transform the star lattice into the honeycomb lattice. The phase transition is the same as the honeycomb lattice where a continuous second-order transition happens.

If $J_{D}$ is antiferromagnetic ($J_{D}<0$), we get the negative critical temperature, which implies no phase transition existing in this case. It can be used as a criterion for experimentalists to determine if the couplings in a real material is ferromagnetic or antiferromagnetic. If one finds a phase transition in the real material, we can conclude that the couplings $J_{D}$ and $J_{T}$ should be both ferromagnetic. In there is no long range order found, it means that at least one kind of nearest neighbor couplings is antiferromagnetic in the material.

%@@@@@@@@@@@@@@@@@@@@@@@@@@@@@@@@@@@@@@@@@@@@@@@@@@@@@@@@@@@@@@@@@@@@@@@@@@@@@@@

\subsection{Partition function}

%%%%%%% F I G U R E %%%%%%%%%%%%
%%%%%%% E N D   F I G U R E %%%%%%%%%%%%

Since we have utilized the $\Delta-Y$ transformation and series reductions to map the star lattice to a honeycomb lattice, the  partition function per unit cell, $z_{s}$, of the star lattice Ising model is equivalent to that of the honeycomb lattice $z_{H}$ multiplied by some coefficients which result from the transformation. These coefficients are as follows,
\begin{align}
&z_{1}=\frac{1}{1+x_{1}^{3}}\sqrt{\frac{x_{1}^3}{x_{T}^3}}\\
z_{2}&=(1+x_{1}x_{D}) \sqrt{\frac{x_{2}}{x_{1}x_{D}}}\\
z_{3}&=(1+x_{1}x_{2})\sqrt{\frac{x_{h}}{x_{1}x_{2}}}
\end{align}
Therefore, the total partition function of the star lattice is
\begin{equation}
z_{s}=z_{1}^2 z_{2}^3 z_{3}^3 z_{H}
\end{equation}
where $z_{H}$ is calculated using the Pfaffian method.\cite{1963JMP4}
We rewrite it here,
\begin{equation}
z_H(x_h)
  =\frac{\sqrt{2} (1 - {x_h}^2)}{x_h}
    \exp \{ \tfrac{1}2 \Omega\left[w(x_h)\right] \}
\end{equation}
where
\begin{equation}
\Omega(w)
 =\int_0^{2\pi} \frac{dp}{2\pi}  \int_0^{2\pi} \frac{dq}{2\pi}
   \ln (w - \cos p - \cos q - \cos (p+q))
\end{equation}
and
\begin{equation}
w(x_h)
  = \frac{1 - 2{x_h} + 6{x_h}^2 - 2{x_h}^3 + {x_h}^4}{2x_h(1-x_h)^2}.
\end{equation}
We can rewrite $\Omega(w)$ and get a more accurate numerical evaluation according to the singularities of the integrand.
\begin{equation}
\Omega(w)
=\frac{2}{\pi} \int_0^{\pi/2} dp
  \ln \left[\cos p + \text{arccosh}  \frac{w-\cos 2p}{2 \cos p}
  \right]
\end{equation}
The partition function of the star lattice Ising model is therefore
\begin{equation}
z_{s}(x_{T},x_{D})
=\Psi(x_{T},x_{D})
        \exp \left[ \tfrac{1}2 \Omega (w (x_h (x_{T}, x_{D}))) \right]
\end{equation}
where
\begin{align}
\Psi(x_{T}&,x_{D})=x_{T}^{-3}x_{D}^{-\frac{3}{2}}(1-x_{T}^2)(1-x_{D}^2)\notag \\
&\times\sqrt{2(1+x_{T}^{2}+2x_{D}x_{T}^{2})(x_{D}+2x_{T}^{2}+x_{D}x_{T}^{2})}
\end{align}
%%%%%%% F I G U R E %%%%%%%%%%%%
\begin{figure}[b]
\centering
{\resizebox*{0.47\textwidth}{!}{\includegraphics{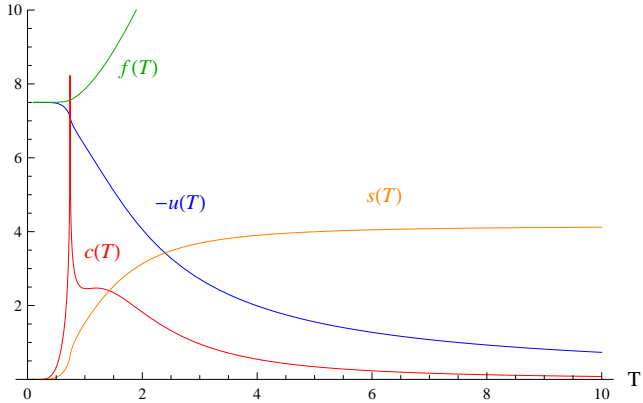}}}
\caption{(Color online). Thermodynamic functions for one unit cell vs temperature $T$ for the unfrustrated case $J_{D}=0.5$ and $J_{T}=1$. The specific heat (red line) diverges as $T\approx 0.74|J_{T}|$ revealing that there is a phase transition from the ferromagnetic phase to the parametric phase. Energy (blue line) is shown as $-u(T)$. Entropy (orange line) approaches zero when $T \rightarrow 0$  and saturates at $6ln2$ as $T\rightarrow \infty$.}
\label{thermodynamicquantity1}
\end{figure}
%%%%%%% E N D   F I G U R E %%%%%%%%%%%%
%%%%%%% F I G U R E %%%%%%%%%%%%
\begin{figure}[t]
\centering
{\resizebox*{0.47\textwidth}{!}{\includegraphics{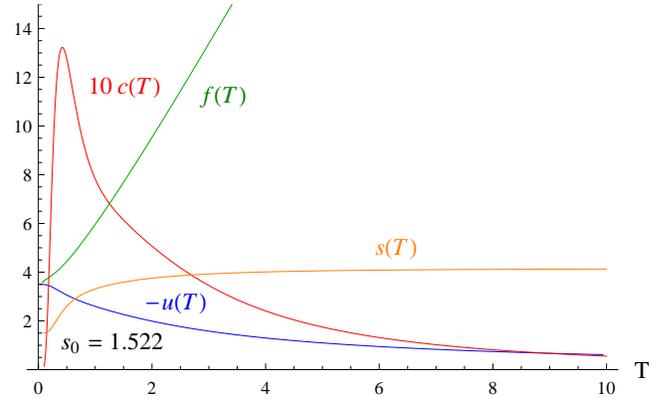}}}
\caption{(Color online). Thermodynamic functions for one unit cell vs temperature $T$ for frustrated coupling $J_{D}=-0.5$ and $J_{T}=-1$. The specific heat (red line) is no longer diverging. Entropy (orange line) remains a none-zero value at $T=0$ and $6ln2$ at high temperature.}
\label{thermodynamicquantity2}
\end{figure}
%%%%%%% E N D   F I G U R E %%%%%%%%%%%%

The total partition function is given by $Z_{s}=z_{s}^{N}$, where $N$ is the spin number of unit cell. Since the partition function is obtained, the internal energy, specific heat, entropy and free energy can be calculated from it.

%%%%%%%%%%%%%%%%%%%%%%%%%%%%%%%%%%%%%%%%%%%%%%%%%%%%%%%%%%%%%%%%%%%%%%%%%%%%%%%
\subsection{Energy}
Taking derivation of the partition function, the energy per unit cell of the star lattice Ising model can be obtained.
\begin{align}
u&=-\frac{d\ln z}{d \beta}
        =
        -\frac{d x_{T}}{d \beta} \frac{\partial \ln z}{\partial x_{T}}
        -\frac{d x_{D}}{d \beta} \frac{\partial \ln z}{\partial x_{D}}\notag
\\
        &=\sum_{i=D,T}
                J_i x_i \left[2\frac{\partial \ln\Psi}{\partial x_i}
                + \frac{\partial x_h}{\partial x_i} \frac{d w}{d x_h} \frac{d \Omega}{d w}
                \right],
        \end{align}
where $\frac{d\Omega}{dw}$ is expressed in terms of the complete elliptic integral of the first kind, K,~\cite{JPSJ.61.64}
 \begin{align}
        \frac{d \Omega}{d w}
               &=-\tfrac2{\pi
   (-w-1)^{3/4}(-w+3)^{1/4}}\notag
\\ \times
               & K
                \left(  \tfrac{1}2+
   \tfrac{w^2-3}{2(w+1)(-w-1)^{1/2}(-w+3)^{1/2}} \right).
        \end{align}
The plots of energy in units of $|J_{D}|$, $\frac{u}{|J_{D}|}$ is illustrated in Figs~\ref{thermodynamicquantity1} and ~\ref{thermodynamicquantity2} for the unfrustrated case and frustrated case respectively.

%%%%%%%%%%%%%%%%%%%%%%%%%%%%%%%%%%%%%%%%%%%%%%%%%%%%%%%%%%%%%%%%%%%%%%%%%%%%%%%

\subsection{Specific heat}
By further derivation, $c=\frac{du}{dT}$, the heat capacity per unit cell can be obtained. The details are shown in Ref.~\onlinecite{ISI:000255457200055}. Here we show the plots of $c$ in Figs~\ref{thermodynamicquantity1} and ~\ref{thermodynamicquantity2}.

In the unfrustrated case, the specific heat $c$ shows a sharp peak at $T\approx 0.74|J_{T}|$ where a phase transition happens. The phase transition point is consistent with the result of Eq.~(\ref{xh}). In addition, there is a broad hump at higher temperature because of the flopping of spins.
Moreover, this hump changes with $R=|\frac{J_D}{J_T}|$. It is obvious when $R<1$ and becomes indistinct when $R=1$. However, it arises again when $R \ge 6$ % This implies that $J_{T}$ is the main interaction in the system. Each unit cell contains 6 $J_T$ and 1 $J_D$ bonds accouting for the ratio 6.
In the unfrustrated case, the sharp peak vanishes which implies no phase transition, consistent with the conclusion drawn from the phase diagram.

%%%%%%%%%%%%%%%%%%%%%%%%%%%%%%%%%%%%%%%%%%%%%%%%%%%%%%%%%%%%%%%%%%%%%%%%%%%%%%%
\subsection{Zero-temperature limit: residual entropy}
The plots of entropy are shown in Figs.~\ref{thermodynamicquantity1} and \ref{thermodynamicquantity2}. Nonetheless, we can expand the partition function in series to gain more information about the residual entropy in the low temperature limit.

In the case of $J_{D}>0$, the partition function can be expanded as
\begin{align}
lnz&=-\frac{3}{2}lnx_{D}-3lnx_{T}+\frac{3}{2}x_{D}^{2}+...\\
u&=-3|J_D|-6|J_T|+6|J_{D}|e^{-4\beta |J_{D}|}+...
\end{align}
The residual entropy is therefore $0$ when $T\rightarrow 0$.
%%%%%%% F I G U R E %%%%%%%%%%%%
\begin{figure}[b]
\centering
{\resizebox*{0.47\textwidth}{!}{\includegraphics{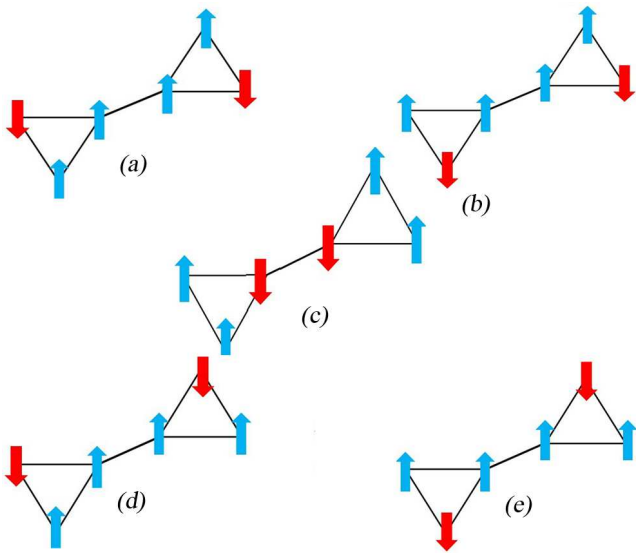}}}
\caption{(Color online). Spin configurations of the degenerate states of phase II. The two spins are coupled by $J_{T}$. Each triangular has exactly two spins pointing up.}
\label{phase2}
\end{figure}
%%%%%%% E N D   F I G U R E %%%%%%%%%%%%
However, when $J_{D}<0$, the model becomes frustrated. In this way, when $T\rightarrow 0$, $\beta \rightarrow \infty$, which means $x_{D}, x_{T}
\rightarrow \infty$, $w \rightarrow \infty$. Therefore, $ln\Omega(w)$ becomes $\sim ln(w)$. Expanding $ln(z)$, we get
\begin{align}
lnz&=\frac{1}{2}ln21 +\frac{3}{2}lnx_{D}+lnx_{T}+...\\
u&=(3|J_{D}+2|J_{T}|)+...
\end{align}
These results contribute to the residual entropy by
\begin{align}
s_0 &= \lim_{\beta J_{T}\rightarrow -\infty} \lim_{\beta
            J_{D}\rightarrow -\infty}
          \left( \ln Z + \beta u\right)=\frac{1}{2}ln21.
          \label{residualentropy}
\end{align}
Thus, the frustration of the system leads to a $\frac{1}{2}ln21 \approx 1.522$ residual entropy per unit cell when $T\rightarrow 0$. One can confirm that, this value is consistent with the entropy at $T=0$ in Fig.~\ref{thermodynamicquantity2}. The residual entropy per site is approximately $0.254$, smaller than the triangle lattice, TKL, and kagome lattice\cite{PhysRev.79.357,ISI:000255457200055,PTP.10.158}. Therefore, the star lattice is less frustrated compared to them. %We have noticed that this value is not a form of $ln n$ where n is the number of degenerate ground states. It will be discussed more detailed in the next section.

%@@@@@@@@@@@@@@@@@@@@@@@@@@@@@@@@@@@@@@@@@@@@@@@@@@@@@@@@@@@@@@@@@@@@@@@@@@@@@@@
\section{Phase diagrams at zero temperature \label{zerotemperature}}
In this section, we present the phase diagrams at zero temperature along with some thermodynamic properties such as energy, magnetization and entropy.
By calculating the ground state energy of the star lattice, we derive the full phase diagram for the system.
Since the phase diagram at zero field is already shown in Fig.~\ref{zerofieldphasediagram}, we focus on the none-zero field case in this section.
The corresponding results are summarized in Figs.~\ref{phasediagram1} and \ref{phasediagram2} according to the sign of $J_T$.
%%%%%%% F I G U R E %%%%%%%%%%%%
\begin{figure}[!t]
\centering
{\resizebox*{0.47\textwidth}{!}{\includegraphics{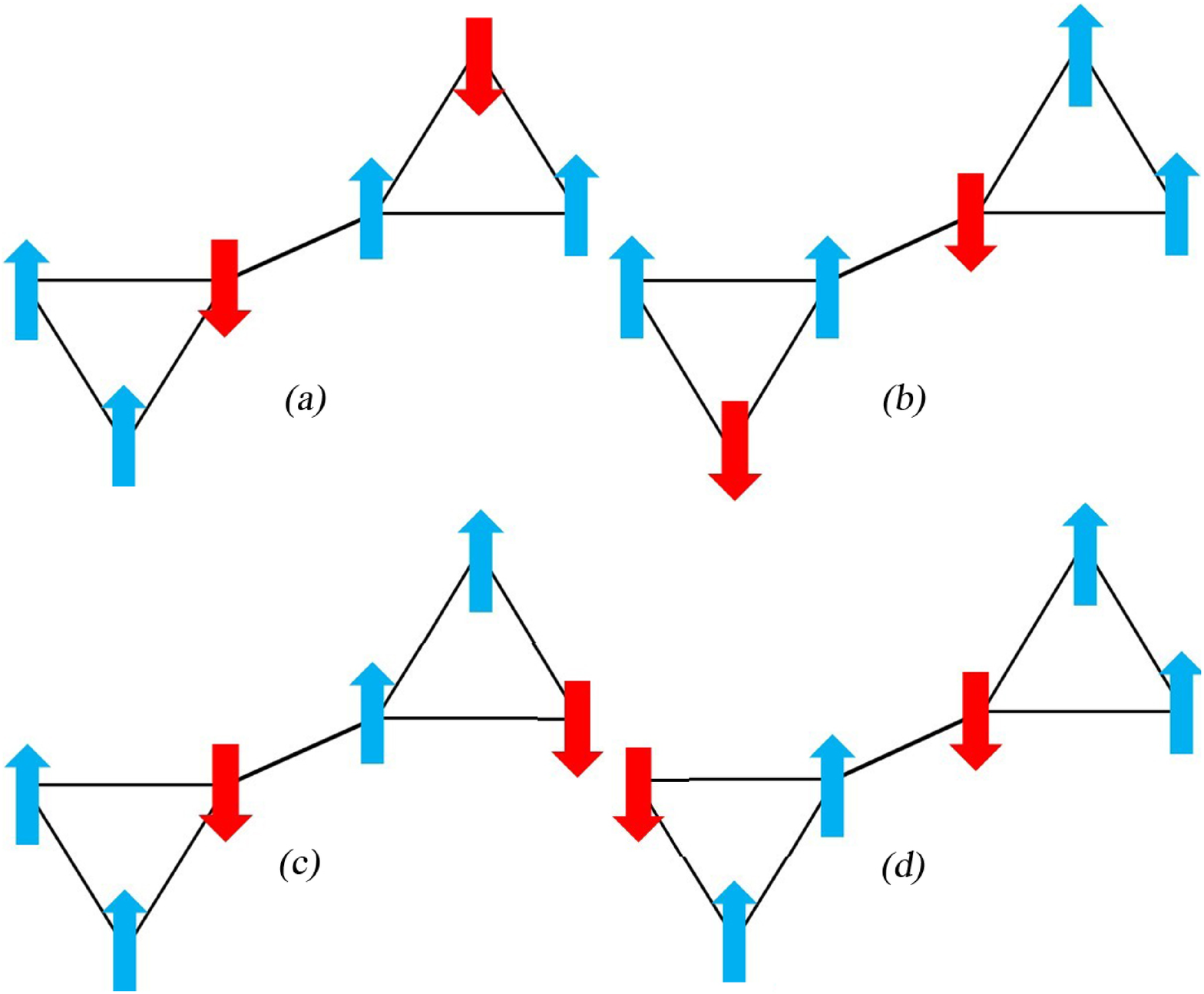}}}
\caption{(Color online). Spin configurations of the degenerate states of phase III. The two spins are coupled by $J_{T}$. Each triangle has exactly two spins pointing up.}
\label{phase3}
\end{figure}
%%%%%%% E N D   F I G U R E %%%%%%%%%%%%

%@@@@@@@@@@@@@@@@@@@@@@@@@@@@@@@@@@@@@@@@@@@@@@@@@@@@@@@@@@@@@@@@@@@@@@@@@@@@@@@
%==============================================================================
\subsection{Zero Field (Phase V and VI)}
The phase diagram for zero field as a function of couplings is showed in Fig.~\ref{zerofieldphasediagram}. When $J_{D}>0$, the phase is ordered and
ferromagnetic. When $J_{D}<0$, the phase is frustrated with a residual entropy $s_{0}=\frac{1}{2}ln21$. When $J_{T}=-1$, the phase is located in the negative section of $T$-axis , which reveals that there is no phase transition in this phase. The disordered and ordered phases are labeled by V and VI in Figs.~\ref{phasediagram1} and ~\ref{phasediagram2} respectively.

In phase V, the system is fully frustrated. We find $18$ degenerate ground states for each unit cell. However, as shown in Eq.~(\ref{residualentropy}), the residual entropy is not $ln18$ but $\frac{1}{2}ln21$. This is similar as the triangular lattice whose residual entropy can not be obtained by counting the number of ground states in a unit cell.~\onlinecite{PhysRev.79.357}
%However, in Ref.~\onlinecite{ISI:000255457200055}, TKL was found with a residual entropy can be interpreted by counting the number of ground states.
%It is interesting to think about questions that why frustrated spin systems behave so differently in residual entropy and what kind of systems has the property like TKL.
%==============================================================================
%%%%%%% F I G U R E %%%%%%%%%%%%
\begin{figure}[t]
\centering
{\resizebox*{0.47\textwidth}{!}{\includegraphics{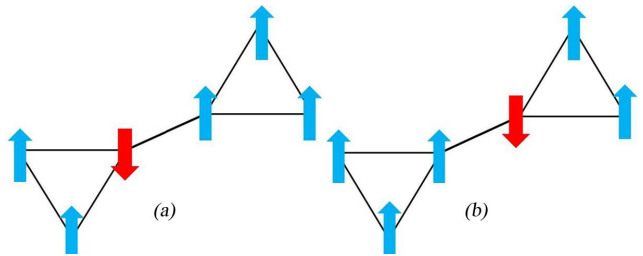}}}
\caption{(Color online). Spin configurations of the degenerate states of phase IV. All the configurations have only one spin points down.}
\label{phase4}
\end{figure}
%%%%%%% E N D   F I G U R E %%%%%%%%%%%%
%%%%%%% F I G U R E %%%%%%%%%%%%
\begin{figure}[t]
\centering
{\resizebox*{0.47\textwidth}{!}{\includegraphics{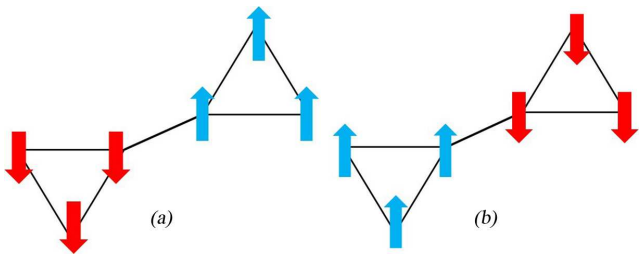}}}
\caption{(Color online). Spin configurations of the degenerate states of phase VII. This is a  interesting phase because spins in the same triangle point the same direction. It can map to a honeycomb lattice with equivalent antiferromagtic coupling with high spins.}
\label{phase7}
\end{figure}
%%%%%%% E N D   F I G U R E %%%%%%%%%%%%
\subsection{Saturated ferromagnetic phase (Phase I)}
When the external field is strong enough, i.e. $h>Max\{2|J_T|,|J_{D}|+2|J_{T}|\}$, the phase is ferromagnetic where all spins are lined up. It is obvious that this state has energy $u=-J_{D}-6J_{T}-6h$, magnetization $m=6$ and entropy $s=0$ per unit cell.

%==============================================================================
\subsection{Phase II}
When the field is weaker, e.g. $0<h<2|J_{T}|$ and $J_{T}<0,J_{D}>0$, it is a phase with $u=-J_{D}+2J_{T}-2h, m=2, s=ln5$. The spin configurations of the degenerate ground states of this phase are shown in Fig.~\ref{phase2}. The two spins connected by $J_D$ become aligned due to the positive $J_D$ and the weak field $h$.

%==============================================================================
%%%%%%% F I G U R E %%%%%%%%%%%%
\begin{figure}[b]
\centering
{\resizebox*{0.47\textwidth}{!}{\includegraphics{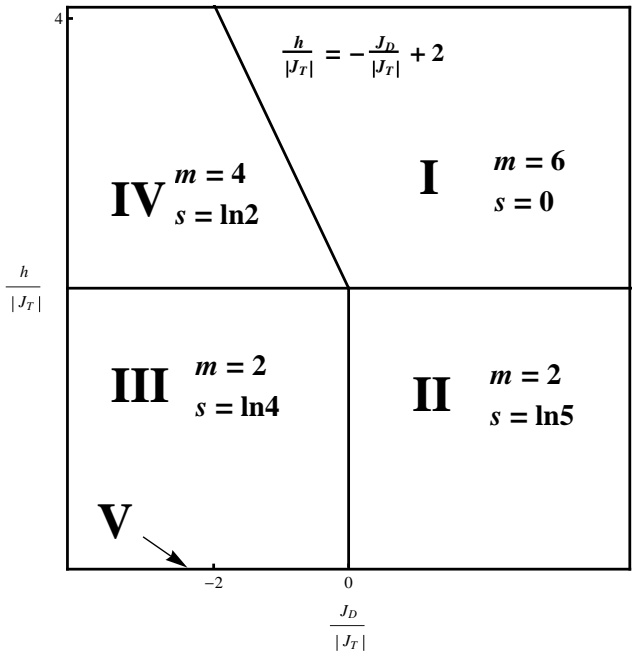}}}
\caption{(Color online). Phase diagram of the star lattice Ising model in the $(J_{D}, h)$ plane for $J_{T}>0$ and $T=0$. The phase diagram is symmetric under a sign change of $h$.}
\label{phasediagram1}
\end{figure}
%%%%%%% E N D   F I G U R E %%%%%%%%%%%%
%%%%%%% F I G U R E %%%%%%%%%%%%
\begin{figure}[b]
\centering
{\resizebox*{0.47\textwidth}{!}{\includegraphics{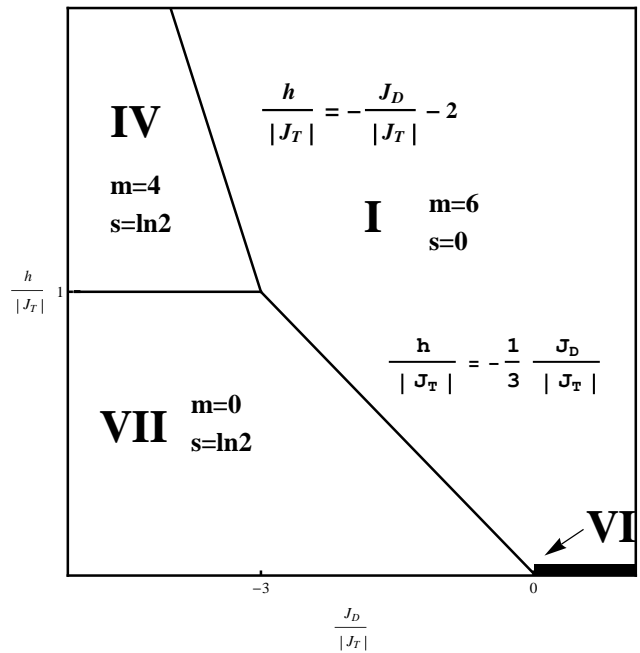}}}
\caption{(Color online). Phase diagram of the star lattice Ising model in the $(J_{D}, h)$ plane for $J_{T}<0$ and $T=0$.}
\label{phasediagram2}
\end{figure}
%%%%%%% E N D   F I G U R E %%%%%%%%%%%%
\subsection{Phase III}
If $0<h<2$, the system is in a frustrated phase. We find four degenerate ground states in this phase contributing to the residual entropy $s=ln4$. The other properties are given by $u=J_{D}+2J_{T}-2h$, and $m=2$. The spin configurations are shown in Fig.~\ref{phase3}. In this case, the two spins connected by $J_T$ become antiparallel since $J_D$ is antiferromagnetic.
%==============================================================================
\subsection{Phase IV}
In the case of $2<h<|J_{D}|+2|J_{T}|$ and $J_{T}>0,J_{D}<0$, phase III evolves into phase IV, which has $m=4$ and $s=ln2$. Only one spin points down in this phase and it should be one of the two connected by $J_D$. The spin configurations are shown in Fig.~\ref{phase4}.
%==============================================================================
\subsection{Phase VII}
Phase VII is a new phase when $J_T$ becomes positive in none-zero field. In this phase, $h<\frac{1}{3}|J_{D}|$ and $J_{D}<0,J_{T}>0$, the spins on the same triangle are parallel, however, antiparallel to the neighboring triangles for the positive $J_T$ and negative $J_D$. If we treat the three spins on the same triangle as a higher spin located in the center of the triangle, it becomes an antiferromagnetic phase in a honeycomb lattice. This state gives $u=J_{D}-6J_{T},s=ln2$ and m=0.

%==============================================================================
\subsection{Phase diagram}
According to the discussion above, we now combine all the results together to obtain a full phase diagram. Fig.~\ref{phasediagram1} show the phase diagram when $J_{T}$ is antiferromagnetic and thus $J_{T}<0$ and Fig.~\ref{phasediagram2}, on the contrary, shows the case when $J_{T}$ is ferromagnetic. The phase diagram is symmetric under the sign change of $h$. %and flipping all the spins over at the same time.

\section{Monte Carlo Simulations \label{montecarlo}}
%%%%%%%%%%%%%%%%%%%%%%%%%%%%%%%%%

In this section we show the Monte Carlo (MC) simulation results of the star lattice Ising model with different combinations of parameters, which helps to corroborate our analytic predictions. Meanwhile, they allow us to calculate the magnetization and susceptibility at finite temperature.

%%%%%%% F I G U R E %%%%%%%%%%%%
\begin{figure}[b!]
\centering
{\resizebox*{0.48\textwidth}{!}{\includegraphics{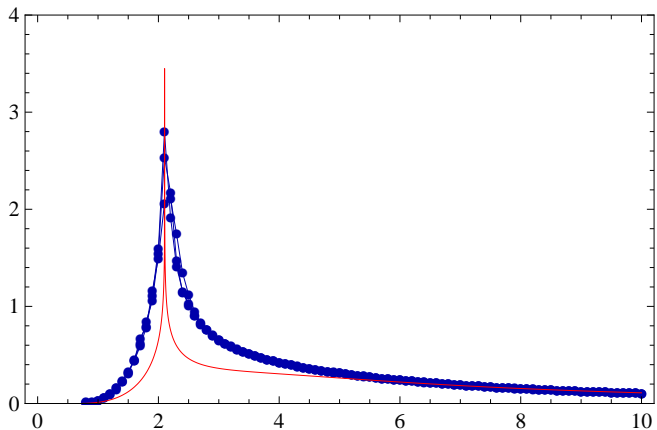}}}
\caption{(Color online). Temperature dependence of heat capacity per site from exact solution (red line) and Monte Carlo simulations with $L=8$, $16$ and $32$ for $J_{D} = 5 J_{T} = 1$. The critical temperature $T_c \approx 2.1$$|J_T|$, consistent with Fig.~\ref{zerofieldphasediagram}.}
\label{mcc}
\end{figure}
%%%%%%% E N D   F I G U R E %%%%%%%%%%%%
%%%%%%% F I G U R E %%%%%%%%%%%%
\begin{figure}[t!]
\centering
{\resizebox*{0.55\textwidth}{!}{\includegraphics{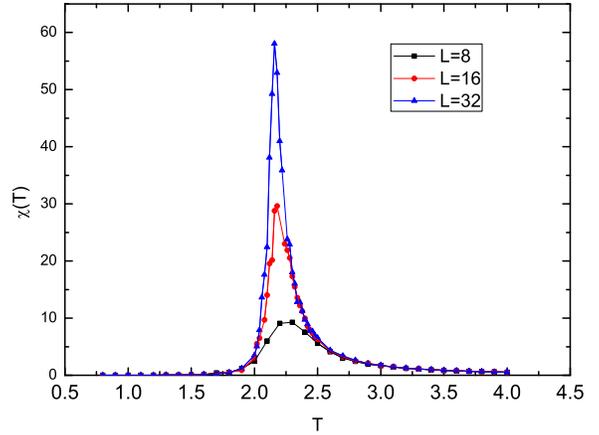}}}
\caption{(Color online). Temperature dependence of susceptibility from the Monte Carlo simulations with $L=8$, $16$ and $32$ for $J_{D} = 5 J_{T} = 1$.}
\label{mcsf}
\end{figure}
%%%%%%% E N D   F I G U R E %%%%%%%%%%%%
%%%%%%% F I G U R E %%%%%%%%%%%%
\begin{figure}[t]
\centering
{\resizebox*{0.55\textwidth}{!}{\includegraphics{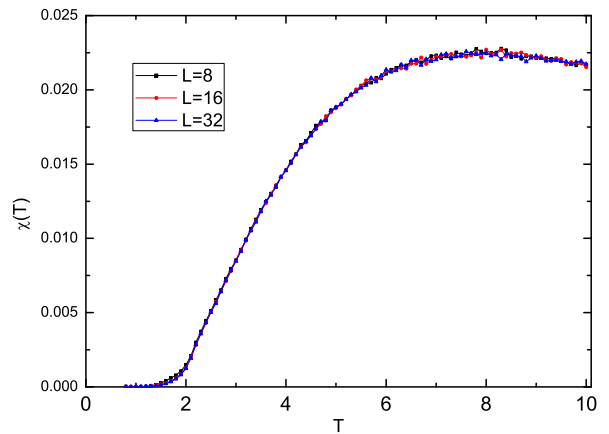}}}
\caption{(Color online). Temperature dependence of susceptibility from the Monte Carlo simulations with $L=8$, $16$ and $32$ for $J_{D} = 5 J_{T} = -1$. There is no apparent peak found, which means there is no continuous phase transition for $J_{T}<0$. The exact solution in Fig.~\ref{zerofieldphasediagram} gives the same result in this case.}
\label{mcsan}
\end{figure}
%%%%%%% E N D   F I G U R E %%%%%%%%%%%%
We choose system size $L=8$, $16$, $32$, where $L$ is the length of the unit cell for the star lattice, which means that the total number of spins $N$ is $N=6L^{2}$, as there are six spins in each unit cell. The periodic boundary condition is used for the simulations.

The specific heat $c$ and magnetic susceptibility $\chi$ during the MC simulations can be calculated using the fluctuation-dissipation theorem
\begin{align}
        c &=\frac{\mean{H^2} - \mean{H}^2}{NT^2},\\
        \chi   &=\frac{\mean{M^2} - \mean{M}^2}{NT},
 \label{chi}
\end{align}
where $\mean{H}$ and $\mean{M}$ are respectively the Monte Carlo averages of the
total energy (i.e., the Hamiltonian) and magnetization.

%%%%%%%%%%%%%%%%%%%%%%%%%%%%%%%%%%%%%%%%%%%%%%%%%%%%%%%%%%%%%%%%%%%%%%%%%%%%%%%

Fig.~\ref{mcc} shows the temperature dependence of heat capacity per site at $h=0$ for a typical unfrustrated case $J_{D}/ J_{T} = 5$, $J_{T} >0$. The MC results are consistent with the exact analytic results.
%The phase transition point and high temperature limit are consistent with the exact results.
%In the Monte Carlo simulations, the peak of heat capacity (Fig.~\ref{mcc}) and

We also calculate the susceptibility from the MC simulations, which is shown in Fig.~\ref{mcsf}. As $L$ increases, the peaks become sharper and sharper, indicating a phase transition.

We also study two different combinations of interactions for the unfrustrated case, which is $J_{T}>0$, in Figs.~\ref{mcsf} and \ref{mcsan}. The temperature dependence of susceptibility is found to be sensitive to the sign of $J_{D}$. When $J_{D}>0$, the susceptibility has a sharp peak at the critical point between the ferromagnetic phase and paramagnetic phase. However, when $J_{D}<0$, there is no such peak, which implies no phase transition in this case, just as what we get in exact solutions. The shape of the susceptibility peak depends on the size of the system when $J_{D}>0$, whereas for $J_{D}<0$, the size of the system has no influence on the susceptibility.
\section{CONCLUSIONS}
In summary, we have studied the Ising model on the star lattice with two different exchange couplings $J_{T}$ and $J_{D}$ using both analytical method and Monte Carlo simulations. We have presented its thermodynamic properties including internal energy, free energy, specific heat, entropy and susceptibility in the zero field. The phase transition temperature for $J_{T}=J_{D}$ is exactly same as the one found in Ref.~\cite{1995PhRvB..51.5840B}.  There is no phase transition found if one of the couplings is antiferromagnetic. Moreover, we have obtained the rich phase diagrams in terms of $J_{T}, J_{D} $ and $h$ at zero temperature. Monte Carlo simulation is used to confirm the exact results and calculate the susceptibility.

In the fully frustrated case, the residual entropy of the system can be expressed as a closed form ($s_{0}=\frac{1}{2}ln21$ as showed in Eq.~(\ref{residualentropy}) which is consistent with the triangular and Kagome lattices. The system is less frustrated compared to the other triangulated lattices.

Our study provides a benchmark calculation for the thermodynamics of Ising spins on the star lattice, which can help experimentalists to investigate the real materials.

\begin{acknowledgments}
We thank Xiao-Ming Chen and Ming-Liang Tong for helpful discussions.
This work is supported by the Fundamental Research Funds for the
Central Universities of China (11lgjc12 and 10lgzd09), NSFC-11074310 and 11275279, MOST of China 973 program (2012CB821400), Specialized Research Fund
for the Doctoral Program of Higher Education (20110171110026),
Undergraduate Training Program at SYSU and NCET-11-0547.
\end{acknowledgments}

\bibliographystyle{apsrev4-1}
%\bibliography{bibiographydata}
%

\end{document}